\documentclass[prc,aps,floatfix,twocolumn,groupedaddress,nofootinbib,showpacs,preprintnumbers,superscriptaddress,
amsmath,amssymb,amsfonts,widetable]{revtex4-2}
\usepackage{array}
\usepackage{hyperref}
\usepackage[nameinlink,capitalize]{cleveref}
\usepackage{xcolor}
\usepackage{graphicx}
\usepackage{bm}
\usepackage{soul}
\usepackage{array}
\usepackage{lipsum}
\usepackage{relsize}
\usepackage{mathrsfs}
\usepackage{amssymb}
\usepackage{amsmath}
\usepackage{graphicx}
\usepackage{microtype}
\usepackage{slashed}

\crefname{section}{Sec.}{Secs.}

\newcommand{\msun}{M_\odot}

\newcommand{\lead}{$^{208}$Pb}
\newcommand{\calfe}{$^{48}$Ca}
\newcommand{\ca}{$^{40}$Ca}

\newcommand{\nn}{\ensuremath{NN}}
\newcommand{\nnn}{\ensuremath{3N}}
\newcommand{\elem}[2]{\ensuremath{^{#2}\text{#1}}}
\newcommand{\hebelerA}{\ensuremath{\text{1.8/2.0~(EM)}}}
\newcommand{\dnnlogo}{\ensuremath{\Delta\text{NNLO}_\text{GO}}}
\newcommand{\nnlosim}{\ensuremath{\text{NNLO}_\text{sim}(550)}}
\newcommand{\arthuisA}{\ensuremath{\text{1.8/2.0~(EM7.5)}}}
\newcommand{\arthuisB}{\ensuremath{\text{1.8/2.0~(sim7.5)}}}
\newcommand{\nnnlo}{\ensuremath{\text{N}^3\text{LO}}}
\newcommand{\nnlo}{\ensuremath{\text{N}^2\text{LO}}}
\newcommand{\fmithree}{\ensuremath{\text{fm}^{-3}}}

\newcommand{\rmfgonodelta}{RMF-${\rm GO}$(no $\delta$)}
\newcommand{\rmfgo}{RMF-${\rm GO}$}


\begin{document}

\title{Connecting Relativistic Density Functional Theory to Microscopic Calculations}

\author{Brendan T. Reed}
\affiliation{
Theoretical Division, Los Alamos National Laboratory, Los Alamos, NM 87545, USA
}
\email{breed@lanl.gov}
\author{Matthias Heinz}
\affiliation{National Center for Computational Sciences, Oak Ridge National Laboratory, Oak Ridge, TN 37831, USA}
\altaffiliation{Notice: This manuscript has been authored by UT-Battelle, LLC, under contract DE-AC05-00OR22725 with the US Department of Energy (DOE). The US government retains and the publisher, by accepting the article for publication, acknowledges that the US government retains a nonexclusive, paid-up, irrevocable, worldwide license to publish or reproduce the published form of this manuscript, or allow others to do so, for US government purposes. DOE will provide public access to these results of federally sponsored research in accordance with the DOE Public Access Plan (\url{https://www.energy.gov/doe-public-access-plan}).}
\affiliation{Physics Division, Oak Ridge National Laboratory, Oak Ridge, TN 37831, USA}
\affiliation{Technische Universit\"at Darmstadt, Department of Physics, 64289 Darmstadt, Germany}
\affiliation{ExtreMe Matter Institute EMMI, GSI Helmholtzzentrum f\"ur Schwerionenforschung GmbH, 64291 Darmstadt, Germany}
\affiliation{Max-Planck-Institut f\"ur Kernphysik, Saupfercheckweg 1, 69117 Heidelberg, Germany}
\author{Pierre Arthuis}
\affiliation{Université Paris-Saclay, CNRS/IN2P3, IJCLab, 91405 Orsay, France}
\affiliation{Technische Universit\"at Darmstadt, Department of Physics, 64289 Darmstadt, Germany}
\affiliation{ExtreMe Matter Institute EMMI, GSI Helmholtzzentrum f\"ur Schwerionenforschung GmbH, 64291 Darmstadt, Germany}
\affiliation{Helmholtz Forschungsakademie Hessen für FAIR (HFHF), GSI Helmholtzzentrum für Schwerionenforschung GmbH, 64291 Darmstadt, Germany}
\author{Achim Schwenk}
\affiliation{Technische Universit\"at Darmstadt, Department of Physics, 64289 Darmstadt, Germany}
\affiliation{ExtreMe Matter Institute EMMI, GSI Helmholtzzentrum f\"ur Schwerionenforschung GmbH, 64291 Darmstadt, Germany}
\affiliation{Max-Planck-Institut f\"ur Kernphysik, Saupfercheckweg 1, 69117 Heidelberg, Germany}
\author{Ingo Tews}
\affiliation{
Theoretical Division, Los Alamos National Laboratory, Los Alamos, NM 87545, USA
}

\begin{abstract}
   The development of systematic effective field theories (EFTs) for nuclear forces and advances in solving the nuclear many-body problem have greatly improved our understanding of dense nuclear matter and the structure of finite nuclei. For global nuclear calculations, density functional theories (DFTs) have been developed to reduce the complexity and computational cost required in describing nuclear systems. 
   However, DFT often makes approximations and assumptions about terms included in the functional, which may introduce systematic uncertainties compared to microscopic calculations using EFTs. 
   In this work, we investigate possible avenues of improving nuclear DFT using nonlinear relativistic mean-field (RMF) theory. 
   We explore the impact of RMF model extensions by fitting the nonlinear RMF model to predictions of nuclear matter and selected closed-shell nuclei using four successful chiral EFT Hamiltonians. 
   We find that these model extensions are impactful and important in capturing the physics present within chiral Hamiltonians, particularly for charge radii and neutron skins of closed-shell nuclei. 
   However, there are additional effects that are not captured within the RMF model, particularly within the isoscalar sector of RMF theory. 
   Additional model extensions and the reliability of the nonlinear RMF model are discussed.
\end{abstract}

\preprint{LA-UR-25-23461, INT-PUB-25-010}
\maketitle

\section{Introduction}

Major advances in low-energy nuclear theory have been made in the past two decades, leading to a systematic microscopic description of nuclei and nuclear matter. 
This was enabled by development of effective field theories (EFTs) for nuclear forces~\cite{Epelbaum2009RMP_ChiralEFTReview, Machleidt2011PR_ChiralEFTReview, Hammer2020RMP_NuclearEFT} combined with the development of powerful many-body methods applicable to systems ranging from light to heavy nuclei and nucleonic matter~\cite{Dickhoff2004PPNP_SCGFReview, Lee2009PPNP_LatticeReview, Barrett2013PPNP_NCSMReview, Hagen2014RPP_CCReview, Carlson2015RMP_QMCReview, Hergert2016PR_IMSRG,Lynn2019ARNPS_QMCReview,Hergert2020FP_AbInitioReview, Drischler2021ARNPS_ChiralEFTEOSReview}. 
The advances in many-body calculations have enabled calculations of some of the heaviest nuclei such as lead~\cite{Miyagi2022PRC_NO2B, Hu2022NP_Pb208, Hebeler2023PRC_JacobiNO2B, Miyagi2024PRL_MagMoments, Door2025PRL_YbBoson, Arthuis2024arxiv_LowResForces} and reliable predictions for the properties of strongly interacting matter in neutron stars~\cite{Hebeler:2009iv,Tews:2012fj,Lynn:2015jua,Drischler2019PRL_MBPTChiralSaturation,Piarulli:2019pfq,Drischler:2020yad,Keller2023PRL_MBPTFiniteTProtonFraction,Tews:2024owl}, both with systematic uncertainty estimates.

These advances also present a unique opportunity to improve other theoretical approaches like density functional theory (DFT). 
DFT is a powerful method for predicting properties of finite nuclei and nucleonic matter due to its low computational cost compared to the aforementioned microscopic approaches. 
While there are many different approaches and parameterizations within DFT, the two main formalisms are nonrelativistic and covariant DFT. 
The widely used Skyrme functional~\cite{Skyrme:1956zz}, a nonrelativistic DFT, has been a popular choice for describing nuclear systems and has a broad range of applicability (see, e.g., Refs.~\cite{Bender2003RMP_DFTReview,Stone:2006fn,Kortelainen:2010hv,Erler2012N_DFTDripLines}. 
The Walecka model~\cite{walecka:1974} is among the most popular covariant (relativistic) DFTs with success in many similar applications. 
This model typically is parameterized in terms of nucleons interacting via the exchange of various scalar/vector mesons with varying degrees of complexity due to nonlinear self-interactions~\cite{Serot_Walecka,Lalazissis:1996rd,Horowitz:2001ya}, density-dependent couplings~\cite{LONG:2006150,Lalazissis:2005de}, or some combination of the two~\cite{Shen:2010pu}.
Additionally, both relativistic and nonrelativistic DFTs utilize the mean-field approximation, which greatly simplifies calculations in both finite nuclei and infinite nuclear matter.
In addition, there have been many studies of neutron stars utilizing RMF theories in their calculations, such as the MUSES project \cite{Dexheimer_2008,ReinkePelicer:2025vuh} or the widely used SFH family of EOSs \cite{Steiner:2012rk}. 

However, DFT approaches do not contain the same physical information as nuclear interactions from chiral EFT (with $\mathcal{O}(20)$ low-energy couplings to be fit to data) due to a lacking microscopic connection to nuclear forces. 
An example is the absence of tensor forces from pion exchange or contact interactions, as this does not contribute to spin-saturated nuclei in the mean-field approximation.
This requires that the effects of the tensor force be absorbed within other effective terms of the functional.

Studies comparing density functionals with microscopic calculations have mainly focused on  neutron and nuclear matter (see, e.g., Refs.~\cite{Goriely:2005job,Typel:2018cap,Marino:2021,alford:2022}). 
In this work, we utilize advances in microscopic calculations with EFT interactions to compare the description of nuclear matter and nuclei from chiral EFT interactions and from a covariant DFT.
Our goal is to identify missing physics and understand areas in which relativistic DFT may be lacking.  
In particular, we use calculations of neutron matter, symmetric nuclear matter, and nuclei up to \elem{Pb}{208} for four chiral EFT interactions as ``pseudo-data'' to adjust new RMF models. 
By comparing the predictions for the different systems, we identify which features in the chiral EFT results can be reproduced by the RMF models. 

This paper is structured as follows.
We provide background information and details of our calculations using both chiral EFT interactions and RMF of finite nuclei and matter in \cref{sec:methods}.
In \cref{ref:fits}, we explain how we fit new RMF models to the results from chiral EFT.
We present our new RMF models in \cref{sec:results}, with results for nuclei and nuclear matter and a comparison of the different models. 
Finally we discuss in \cref{sec:discussion} the implications of using chiral EFT predictions in calibration of RMF models and discuss possible extensions or refinements that can be obtained within the RMF theory used here.

\section{Approach}
\label{sec:methods}

\subsection{Relativistic mean-field theory}

In this work, we use the FSUGold-like class (also called nonlinear) of RMF models \cite{FSUGold,Mueller:1996pm,Horowitz:2000xj,Lalazissis:1999}.
The interaction Lagrangian for this class of models is given as
\begin{align}
    \nonumber \mathcal{L}_{\mathrm{int}}^0&= \bar{\psi}\Big[g_s\phi-\big(g_vW_\mu+\frac{g_\rho}{2}\bm{\tau}\cdot \bm{b_\mu}+\frac{e}{2}(1+\tau_3)A_\mu\big)\gamma^\mu\Big]\psi\\
    &\quad-\frac{\kappa}{3!}(g_s\phi)^3-\frac{\lambda}{4!}(g_s\phi)^4+\frac{\zeta}{4!}g_v^4(W_\mu W^\mu)^2 \nonumber\\
    &\quad+\Lambda_v(g_v^2W_\mu W^\mu)(g_\rho^2\bm{b_\mu}\cdot\bm{b^\mu})\,,
    \label{eq:lagrangian}
\end{align}
wherein nucleon ($\psi$) fields interact via meson exchange through scalar $\sigma$ meson ($\phi$), vector $\omega$ meson ($W_\mu$), and the vector-isovector $\rho$~meson ($\textbf{b}_\mu$) fields. 
Additionally, this Lagrangian contains several self- and cross-interaction terms among the meson fields. 
The cubic and quartic scalar interaction terms ($\kappa$ and $\lambda$) have been shown to be important for reducing the incompressibility of nuclear matter. 
The quartic vector self-coupling $\zeta$ is important for softening the high-density equation of state (EOS) and is particularly important for reducing the maximum mass of neutron stars.
Finally, the $\omega$--$\rho$ cross-coupling term $\Lambda_v$ is important in softening the density dependence of the symmetry energy.

Nonlinear RMF models have been a very powerful tool in describing a number of nuclear systems~\cite{walecka:1974,horowitz:1981,Lalazissis:1999,Glendenning:2000,Carrier:2003,Piekarewicz:2001nm}. 
For many years, the isovector sector of this class of models was sufficient to explain the then-known properties of nuclear systems. 
Accurate fits of nonlinear RMF models have usually included several properties of nuclei including nuclear masses and charge radii \cite{Fattoyev:2010mx,FSUGold2}. 
However, this class of models has also been shown to have difficulties in reconciling constraints from microscopic calculations and isovector radius measurements.
In particular, nonlinear RMF models tend to predict stiff equations of state, which in turn predict, for example, larger neutron skins in nuclei, larger radii for neutron stars, and stiffer behavior of the density dependence of the symmetry energy \cite{big_apple,Lalazissis:1999,Reed:2023}.

The PREX and CREX parity-violating electron scattering experiments have been able to constrain the neutron skins in both $^{208}$Pb and $^{48}$Ca, respectively, which have set a large spark in the DFT community as reconciliation of both experiments has thus far proven to be difficult with current DFT knowledge \cite{Zhang:2022bni,Papakonstantinou:2022,Reinhard:2022inh,Reed:2023,salinas:2024}.
Indeed, the CREX analysis favored a small neutron skin, too small to be reproduced by any FSUGold-type RMF model without sacrificing fundamental nuclear properties.
However, since the conclusion of the experiments, it has been suggested that a strong isovector spin-orbit force may be able to reconcile the discrepancy \cite{Yue:2024,Zhao:2024,Kunjipurayil:2025xss}.
In addition to neutron-skin measurements, the inference of the tidal deformability of a 1.4$\msun$ neutron star, $\Lambda_{1.4}$, from the gravitational-wave event GW170817 \cite{Abbott:PRL2017,Abbott:2018exr}, which saw the merger of two neutron stars, and subsequent analyses favor soft EOSs. 
It was quickly reported after the PREX-2 experiment was conducted that there may be tension between $\Lambda_{1.4}$ and the neutron skin of $^{208}$Pb within this class of nuclear models~\cite{Reed_2021p2}.

Some recent works \cite{salinas:2024,Kunjipurayil:2025xss} have explored the addition of terms to the Yukawa sector of the Lagrangian and their impact on finite nuclei and neutron stars. 
Three new terms were added to the FSUGold-like Lagrangian, which have a noticeable impact on finite nuclei:
\begin{eqnarray}
    \mathcal{L}_{\mathrm{int}}^{\rm ext}=\bar{\psi}\Big[\frac{g_\delta}{2}\bm{\tau}\cdot\bm{\delta}-f_v\frac{\sigma^{\mu\nu}}{2M}\partial_\nu W_\mu-f_\rho\frac{\sigma^{\mu\nu}}{2M}\partial_\nu\bm{b}_\mu\Big]\psi\,.
    \label{eq:extensions}
\end{eqnarray}
These additional terms include the vector-isoscalar $\delta$~meson ($\bm{\delta}$) and tensor couplings $f_v$ and $f_\rho$, which couple to the derivatives of the $\omega$ and $\rho$ fields, respectively \cite{rufa:1988}. 
The addition of the $\delta$~meson provides a simple yet powerful ability to reduce the density dependence of the symmetry energy \cite{Reed:2023}. 
The $\omega$~tensor coupling $f_v$ is sensitive to changes in the $\omega$~meson field and particularly changes in the nuclear surface. 
Because of the importance of the $\omega$~meson in the calculation of nuclear masses, this coupling is also important in adjusting the binding energy of nuclei.
The $\rho$~tensor coupling $f_\rho$ has a significant impact on the charge and weak density of nuclei but is also sensitive to surface effects. 
Recently it has been shown that in the nonrelativistic limit of RMF theory, these tensor couplings directly contribute to the spin-orbit splitting in nuclei \cite{Kunjipurayil:2025xss}.
As a result, both of their inclusions have a larger impact on the predictions of lighter nuclei such as $^{48}$Ca than on heavier nuclei such as $^{208}$Pb.

\subsection{Microscopic calculations}

A microscopic description of nuclei and nuclear matter is possible using \textit{ab initio} methods~\cite{Lynn2019ARNPS_QMCReview,Hergert2020FP_AbInitioReview, Drischler2021ARNPS_ChiralEFTEOSReview} that solve the many-body Schrödinger equation employing nucleon-nucleon (\nn{}) and three-nucleon (\nnn{}) forces from chiral EFT rooted in quantum chromodynamics~\cite{Epelbaum2009RMP_ChiralEFTReview, Machleidt2011PR_ChiralEFTReview,Hammer2020RMP_NuclearEFT}.
Recent progress has enabled microscopic calculations of nuclei as heavy as \elem{Pb}{208}~\cite{Miyagi2022PRC_NO2B, Hu2022NP_Pb208, Hebeler2023PRC_JacobiNO2B, Door2025PRL_YbBoson, Arthuis2024arxiv_LowResForces}, and we leverage this to fit new RMF models to calculations of pure neutron matter, symmetric nuclear matter, and a broad range of closed (sub-)shell nuclei up to \elem{Pb}{208} consistently employing the same input Hamiltonians~\cite{Arthuis2024arxiv_LowResForces,Alp2025_NuclearMatterMBPT}.
This consistency is key, as the nuclear forces employed can be validated against nucleon-nucleon scattering and properties of few-body systems, heavier nuclei, and nuclear matter without any phenomenological tuning or uncontrolled extrapolations.

We consider four Hamiltonians from chiral EFT, the \hebelerA{}~\cite{Hebeler2011PRC_SRG3NFits}, \dnnlogo{}~\cite{Jiang2020PRC_DN2LOGO}, \arthuisA{}, and \arthuisB{} Hamiltonians~\cite{Arthuis2024arxiv_LowResForces}, which differ in their construction and how they are fit to data.
We highlight the main differences between the Hamiltonians and their known strengths and systematic deficiencies
and refer to Refs.~\cite{Hebeler2011PRC_SRG3NFits, Jiang2020PRC_DN2LOGO, Arthuis2024arxiv_LowResForces} for further details.
The \hebelerA{} Hamiltonian~\cite{Hebeler2011PRC_SRG3NFits} is constructed using the \nn{} potential at next-to-next-to-next-to-leading order (\nnnlo{}) from Ref.~\cite{Entem2003PRC_EM500} transformed to lower resolution using similarity renormalization group techniques~\cite{Bogner2010PPNP_RGReview}.
The \nnn{} potential is taken at next-to-next-to-leading order (\nnlo{}) with a relatively low regularization scale of 2.0~fm$^{-1}$ (394~MeV), and the two short-range low-energy couplings (LECs) are fit to the energy of \elem{H}{3} and the radius of \elem{He}{4}.
The \hebelerA{} Hamiltonian has been very successful in nuclear structure, especially for predictions of ground-state energies and excitation spectra~\cite{Hagen2016PRL_Ni78, Stroberg2017PRL_VSIMSRG, Simonis2017PRC_ChiralSaturationNuclei, Stroberg2019ARNPS_VSIMSRG, Stroberg2021PRL_AbInitioLimits}, but it substantially underpredicts experimental charge radii in medium-mass and heavy nuclei due to a nuclear matter saturation point at too high of a density~\cite{Simonis2017PRC_ChiralSaturationNuclei, Drischler2019PRL_MBPTChiralSaturation, Alp2025_NuclearMatterMBPT}.

A simultaneous description of nuclear ground-state energies, charge radii, and nuclear matter properties is challenging using chiral EFT Hamiltonians~\cite{Ekstrom2015PRC_N2LOsat, Hoppe2019PRC_ChiralMedMass, Huther2020PLB_EMNSRG},
but the \dnnlogo{}, \arthuisA{}, and \arthuisB{} Hamiltonians deliver substantially improved descriptions of these observables, achieved through targeted optimization to properties of medium-mass nuclei and/or nuclear matter.
The \dnnlogo{} Hamiltonian~\cite{Jiang2020PRC_DN2LOGO} is constructed with \nn{} and \nnn{} potentials at \nnlo{} with a regularization scale of 394~MeV and with explicit inclusion of $\Delta$ isobars in the EFT.
It is fit to $NN$ scattering data, few-body systems, and nuclear matter properties and additionally optimized to ground-state properties of medium-mass nuclei.
As a result, it is generally successful in the simultaneous description of ground-state energies and charge radii of medium-mass and heavy nuclei~\cite{Jiang2020PRC_DN2LOGO, Arthuis2024arxiv_LowResForces}.
The \arthuisA{} and \arthuisB{} Hamiltonians employ the same construction as the \hebelerA{} Hamiltonian, with \arthuisB{} being constructed using the \nnlosim{} potential of Ref.~\cite{Carlsson2016PRX_NNLOsim}, but using a modified fit of the \nnn{} LECs to the properties of \elem{H}{3} and, crucially, \elem{O}{16}~\cite{Arthuis2024arxiv_LowResForces}.
The optimization to simultaneously reproduce the ground-state energy and charge radius of \elem{O}{16} results in larger values for the \nnn{} LEC $c_D$, yielding a very good reproduction of ground-state energies and charge radii across the nuclear chart.

For nuclear structure properties, we consider ground-state energies, charge radii, and neutron skins for \elem{O}{16,24}, \elem{Ca}{40,48}, \elem{Ni}{56,68,78}, \elem{Sn}{100,108,120,132}, and \elem{Pb}{208}.
These were computed in Ref.~\cite{Arthuis2024arxiv_LowResForces} based on the four aforementioned Hamiltonians using the in-medium similarity renormalization group (IMSRG)~\cite{Tsukiyama2011PRL_IMSRG, Hergert2016PR_IMSRG}.
The IMSRG solves the many-body Schrödinger equation through a continuous unitary transformation of the Hamiltonian that decouples the ground state from its excitations.
It is typically truncated at the level of normal-ordered two-body operators, the IMSRG(2), an approximation that has been demonstrated to be reliable for ground-state properties of medium-mass nuclei~\cite{Heinz2021PRC_IMSRG3, Heinz2025PRC_CaIMSRG3}.
Charge radii are computed using the consistent transformation of translationally invariant point-proton and spin-orbit radius operators, including corrections based on proton and neutron charge radii and the relativistic Darwin-Foldy term~\cite{Friar1997PRA_DarwinFoldy, Ong2010PRC_SpinOrbit, Hagen2016NP_Ca48Skin} (for more details see Ref.~\cite{Heinz2025PRC_CaIMSRG3}).
Neutron skins are computed as the difference of the point-neutron and point-proton radii.

We also consider the energy per particle of symmetric and pure neutron matter.
These were computed in Ref.~\cite{Alp2025_NuclearMatterMBPT} using many-body perturbation theory (MBPT) for particle number densities $\rho = 0.04$--$0.32~\fmithree$~\cite{Hebeler2010PRC_ChiralEFTNeutronMatter, Tews2013PRL_MBPTNeutronMatter, Drischler2019PRL_MBPTChiralSaturation,Keller2023PRL_MBPTFiniteTProtonFraction}.
MBPT (along with other microscopic methods for nuclear matter~\cite{Dickhoff2004PPNP_SCGFReview,Hagen:2013yba,Carbone:2014mja,Carlson2015RMP_QMCReview,Lynn2019ARNPS_QMCReview, Marino2024PRC_ADC3Matter}) has been established as a robust approach to compute nuclear-matter properties up to and beyond nuclear saturation density.
This gives direct constraints for the equation of state in the inner crust and outer core of neutron stars~\cite{Hebeler2013AJ_EOSPP, Greif2020AJ_EOSConstraints, Keller2023PRL_MBPTFiniteTProtonFraction, Keller2024PRL_ProtonDrip}, which both directly constrain the structure of neutron stars and aid in the inference of the equation of state from multi-messenger observations~\cite{Raaijmakers:2021uju,Huth2022N_HICEOS,Pang:2022rzc, Rutherford2024AJL_NICEREOS,Somasundaram:2024ykk}.
In this work, MBPT diagrams involving \nn{} and normal-ordered \nnn{} potentials are included up to third order
while diagrams involving residual \nnn{} potentials are included up to second order.

\textit{Ab initio} calculations rely on systematically improvable approximations for both nuclear Hamiltonians (due to truncation at a finite order in the EFT) and the solution of the many-body Schrödinger equation in medium-mass and heavy nuclei and nuclear matter (through the use of many-body expansion or statistical methods).
As a result, predictions have intrinsic theory uncertainties, which can be quantified through studies of the order-by-order EFT convergence and the convergence of the many-body approach~\cite{Epelbaum2015EPJA_EKMN3LO, Furnstahl2015JPGNPP_EFTBayesianUQ, Wesolowski2016JPGNPP_BayesianInferenceEFT, Hagen2016PRL_Ni78, Simonis2019EPJA_CCEMResponses, Drischler:2020yad,Heinz2021PRC_IMSRG3, Heinz2025PRC_CaIMSRG3,Armstrong:2025tza,Somasundaram:2024ykk}.
In this work, we explore the Hamiltonian uncertainties through the use of the four Hamiltonians detailed above.
While such a selection is not fully representative, the differences between Hamiltonians give us some insight into the effects of the underlying uncertainties.

Uncertainties due to the use of approximate many-body methods are challenging to assess, as systematically exploring these requires using higher-order, higher-precision many-body methods, which are numerically very expensive and also formally challenging to develop~\cite{Hagen2007PRC_CC3N, Binder2013PRC_LambdaCCSDT3N, Soma2013PRC_GorkovSCGF, Heinz2021PRC_IMSRG3}.
We assess the uncertainties for our IMSRG(2) predictions~\cite{Arthuis2024arxiv_LowResForces} based on estimates from Ref.~\cite{Heinz2025PRC_CaIMSRG3} from IMSRG(3)-$N^7$ calculations, which are the next order in the many-body expansion for the IMSRG, and informed by previous studies comparing the IMSRG(2) to other many-body methods~\cite{Hergert2020FP_AbInitioReview}.
We use the following estimates: 
ground-state energies have a 2\,\% uncertainty on the correlation energy, the energy difference between the IMSRG(2) ground-state energy and the Hartree-Fock energy;
charge radii have a 1\,\% uncertainty;
and neutron skins have a 5\,\% uncertainty.
We further estimate the uncertainties for our MBPT predictions for nuclear matter as the difference between second- and third-order predictions at each density.

\section{Fitting RMF models to microscopic calculations}
\label{ref:fits}

We fit our RMF models to properties of nuclei and nuclear matter from chiral EFT calculations using the Hamiltonians described above.
We note that these fits do not include the experimental values for finite nuclei included in our fit, as we are purely trying to find missing physics in the RMF models when compared to the chiral Hamiltonians.
Here, we detail the fitting procedure used to construct our RMF models. 
We first explore the effectiveness of including the model extensions proposed in \cref{eq:extensions} by performing two fits to the results for the $\Delta$NNLO$_{\rm GO}$ interaction:
The first fit is based on a model that follows the Lagrangian in \cref{eq:lagrangian} and a second fit includes the proposed model extensions.
We refer to these models as \rmfgonodelta{} and \rmfgo{}, respectively. 
For all other fits, we use the full Lagrangian including the model extensions.
For the full Lagrangian, our set of free parameters is
\begin{equation}
    \theta = \{m_s, g^2_s , g^2_v, g^2_\rho , g^2_\delta , \kappa, \lambda, \zeta , \Lambda_v, f_v , f_\rho \}\,,
\end{equation}
wherein we fix the $\omega$, $\rho$, and $\mathbf{\delta}$ meson masses at their experimental values \cite{PDG2020}. 
To optimize the parameter set for each model, we utilize the Levenberg-Marquardt (LM) algorithm \cite{curve_fit}. 

Previous studies in a similar approach \cite{hornick,alford:2022} have performed fits to nuclear matter calculations alone. 
These studies mainly focus on the impact of microscopic calculations on the structure of neutron stars by fitting RMF interactions to the PNM and SNM EOSs near saturation density. 
We will follow a similar approach by also fitting to the Hamiltonian predictions of the energy per particle of infinite nuclear matter. 
In order to avoid overfitting to nuclear matter, we tabulate the SNM and PNM EOSs at densities in the range of 0.10--0.22 fm$^{-3}$ in steps of 0.02 fm$^{-3}$.
This ensures equal weights in the least-squares fitting procedure. 
For error estimation on each point, we take the difference between second- and third-order corrections to the energy per nucleon. 

\begin{figure*}
    \centering
    \includegraphics[width=\linewidth]{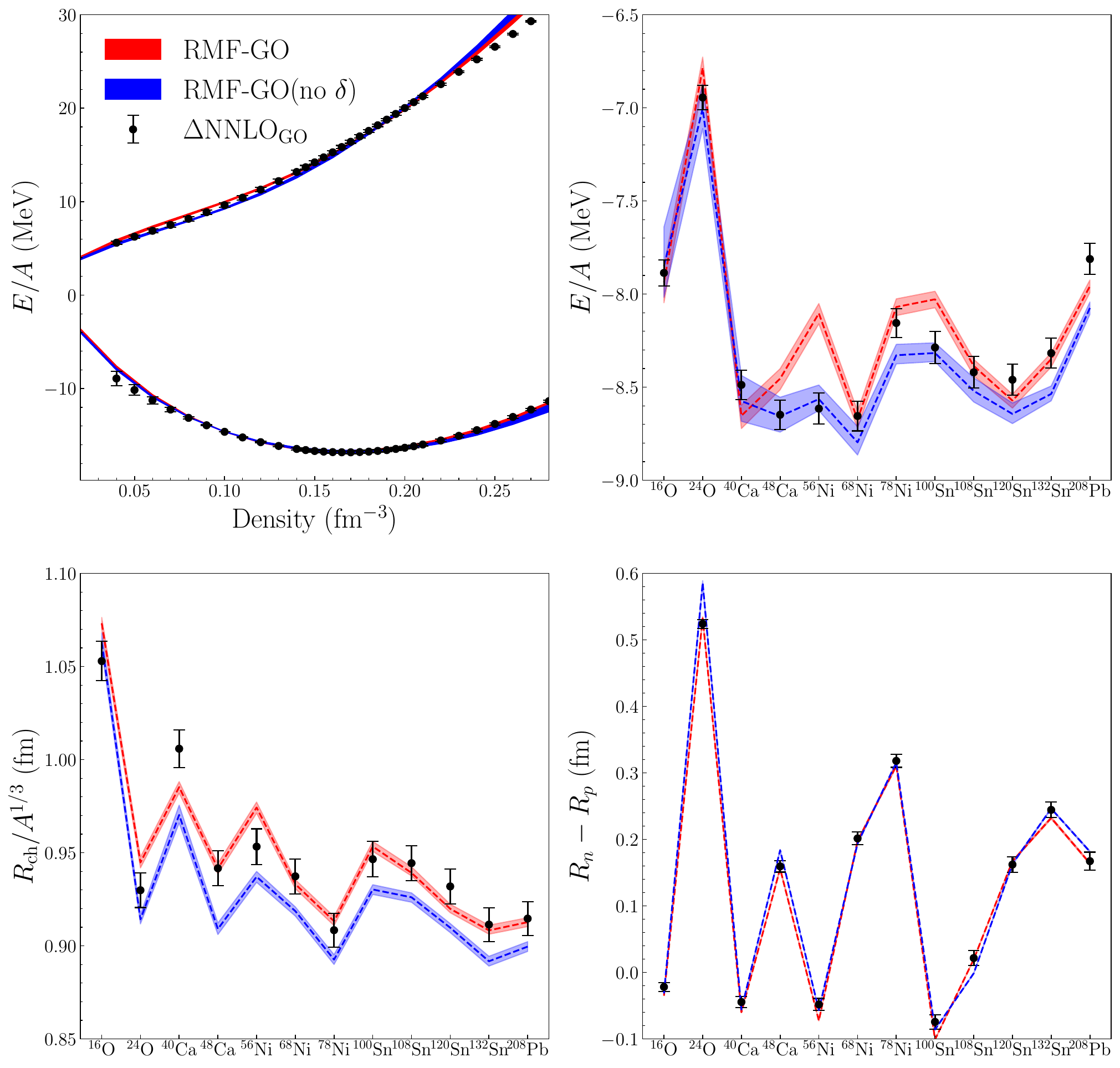}
\caption{Comparison of \rmfgonodelta{}~(blue) and \rmfgo{}~(red) fits (with bands indicating 68\,\% confidence intervals arising from the fit) to the predictions of \dnnlogo{} (black) for selected observables for nuclear matter and nuclei from Refs.~\cite{Arthuis2024arxiv_LowResForces,Alp2025_NuclearMatterMBPT}: (a) energy per nucleon of SNM and PNM; (b) energy per nucleon of closed-shell nuclei; (c) charge radii scaled by $A^{1/3}$; (d) neutron skins ($R_n-R_p$).}
\label{fig:RMFGO}
\end{figure*}

In order to fully determine the RMF Lagrangian, the fits to nuclear matter must be supplemented by fits to finite nuclei. 
Previous studies suggest that the optimal target nuclei for fitting are $^{40}$Ca, $^{48}$Ca, and $^{208}$Pb \cite{Serot_Walecka,big_apple,FSUGold2,Reed:2023}. 
For our RMF calculations of finite nuclei, we use the Hartree formalism based on Ref.~\cite{horowitz:1981}. 
To further constrain and obtain information on the short-range interactions of the scalar meson, we include the energy of each of the three aforementioned nuclei in our fit. 
Additionally, we include the charge radius of $^{40}$Ca and the neutron skins ($R_n-R_p$) of $^{48}$Ca and $^{208}$Pb.\footnote{We have tried several different techniques and target nuclei to fit our models, and this method produced the best results.} 
Specifically, the charge radius of $^{40}$Ca is particularly sensitive to spin-orbit forces, which the tensor couplings are directly linked to \cite{salinas:2024,Kunjipurayil:2025xss}. 
Finally, to explore the uncertainties of the fits on observables, we perform a Markov chain Monte Carlo (MCMC) sampling centered at the best-fit results from the LM algorithm.
The likelihood for the MCMC sampling is constructed from the chiral EFT points whose 1$\sigma$ errors we take as discussed in the previous section. The log-likelihood is explicitly defined as
\begin{equation}
    \log\mathscr{L}=-\frac{1}{2}\sum\frac{y_{\chi}-\hat{y}(\theta)}{\sigma_\chi}\,,
\end{equation}
where subscript $\chi$ represents data from the chiral  Hamiltonians and $\hat{y}$ is the vector of data obtained from the RMF model using parameter vector $\theta$ defined previously. All subsequent uncertainties pertaining to the RMF models will be quoting the 68\,\% uncertainties from the MCMC sampling.

\section{Results}
\label{sec:results}

\subsection{Impact of model extensions}

We first explore how the model extensions change how well the RMF models can simultaneously reproduce the chiral EFT predictions for nuclei and nuclear matter. 
In \cref{fig:RMFGO}, we compare the \rmfgo{} and \rmfgonodelta{} models by considering their predictions for the energy per particle of PNM and SNM and the energies, charge radii, and neutron skins of a broad range of closed-shell nuclei.
For comparison, we also show the predictions from \textit{ab initio} calculations using the \dnnlogo{} interaction to which both models were fit.
The lines for each model are the model predictions, and the bands are the 68\,\% uncertainties based on the uncertainties in the fit that result from the input data.

Both models perform quite similarly for nuclear matter with only modest differences occurring at higher densities where the $\delta$~meson softens the PNM EOS.
This is as expected based on the nature of the $\delta$~meson, which softens the symmetry energy and lowers the effective neutron mass.
What is more apparent are the differences in predictions for finite nuclei.
First, we find that the energy per nucleon in the nuclei considered here show a strong difference between the two fits especially for $^{48}$Ca, $^{56,78}$Ni, $^{100}$Sn, and the heavier nuclei. 
For these medium-mass nuclei the model extensions result in a worse reproduction of the \textit{ab initio} predictions, while the opposite is true for heavier nuclei. 
There is also an interesting underbinding feature found at $^{56}$Ni and $^{100}$Sn in \rmfgo{} compared to \rmfgonodelta{}, which is contrasted by an overbinding in $\elem{Ca}{40}$. 
This is suggestive of some remaining deficiencies in the isoscalar sector of the RMF Lagrangian.
We note that the predicted uncertainties of our RMF models are consistent with the input \textit{ab initio} uncertainties for the energies per nucleon of nuclear matter and nuclei, but considerably smaller than the input uncertainties for charge radii and neutron skins.
We understand this to be due to the strong constraints imposed on our RMF functionals by the fits to nuclear matter.
Within these constraints the RMF models do not have the freedom to full reproduce the input uncertainties for radius observables, leading to smaller predicted uncertainties from the RMF models.

However, we do find that the model extensions greatly improve the reproduction of charge radii. 
The predictions of \rmfgo{} are much closer to the predictions of \dnnlogo{} except for \elem{O}{24}, \elem{Ca}{40}, and \elem{Ni}{56}.
The simultaneous reproduction of the nearly equal charge radii of \elem{Ca}{40} and \elem{Ca}{48} is a well-known challenge for RMF theory, which we believe largely makes up the disagreement we see here.
It is possible that the addition of Fock terms could help alleviate this issue, but this is mostly unstudied for the nonlinear RMF approach \cite{Horowitz:1983,LONG:2006150}.
Therefore, it would be interesting to revisit this issue in the future with the inclusion of Fock terms. 
However, we conclude that the extensions are important in reproducing the charge radii of closed-shell nuclei with $A>56$. 
Finally, we also find that the neutron skins are well reproduced by both models with a slight improvement through the model extensions in \rmfgo{}.

\begin{figure*}
    \centering
    \includegraphics[width=0.8\linewidth]{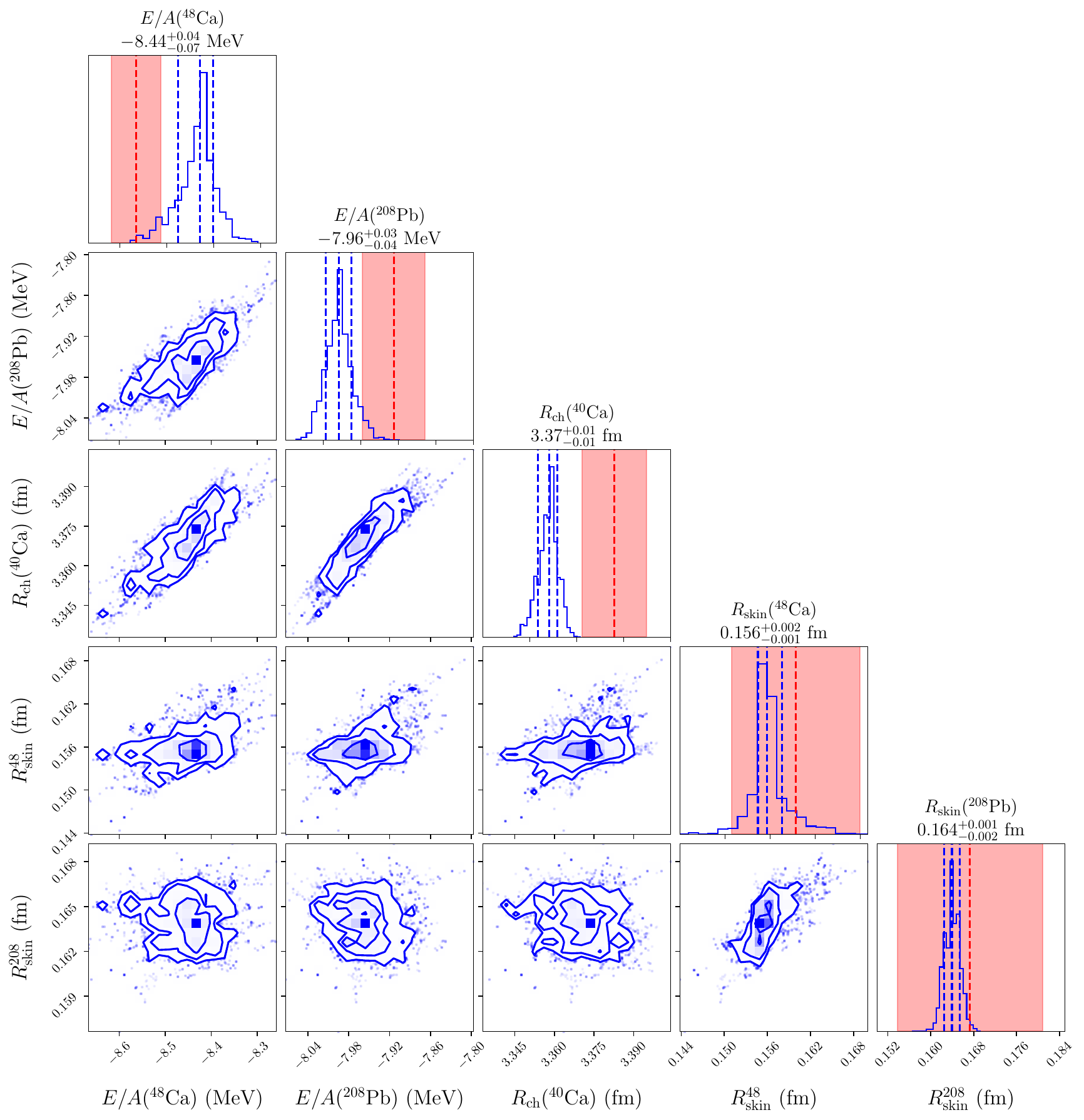}
    \caption{MCMC posterior sample distributions of selected nuclear observables for the \rmfgo{} model: energy per particle $E/A$ of \elem{Ca}{48} and \elem{Pb}{208}; charge radius $R_\mathrm{ch}$ of \elem{Ca}{40}; neutron skin of \elem{Ca}{48} ($R_\mathrm{skin}^{48}$);
    and neutron skin of \elem{Pb}{208} ($R_\mathrm{skin}^{208}$). The 1D histograms show the posterior distributions from the MCMC sampling in blue with the mean and 68\,\% confidence intervals as vertical dashed lines. We also show red bands indicating the assessed uncertainties for the \textit{ab initio} predictions using the \dnnlogo{} Hamiltonian with the central values as red dashed lines.}
    \label{fig:posterior}
\end{figure*}

In \cref{fig:posterior}, we show the MCMC samples of a few nuclear structure observables for the \rmfgo{} model. 
We see that within this model, there is a strong correlation between the charge radius of \ca\ and the energies of \calfe\ and \lead, further suggesting that deficiencies in the isoscalar sector of the Lagrangian are the dominant sources of error in reproducing the chiral EFT results of finite nuclei.
By comparison, the neutron skins of \calfe\ and \lead\ show very weak correlation with the other isoscalar observables and are both consistent with the chiral EFT results, indicating that the isovector sector is currently sufficient to reproduce isovector observables.

\begin{figure*}[htb]
    \includegraphics[width=.9\linewidth]{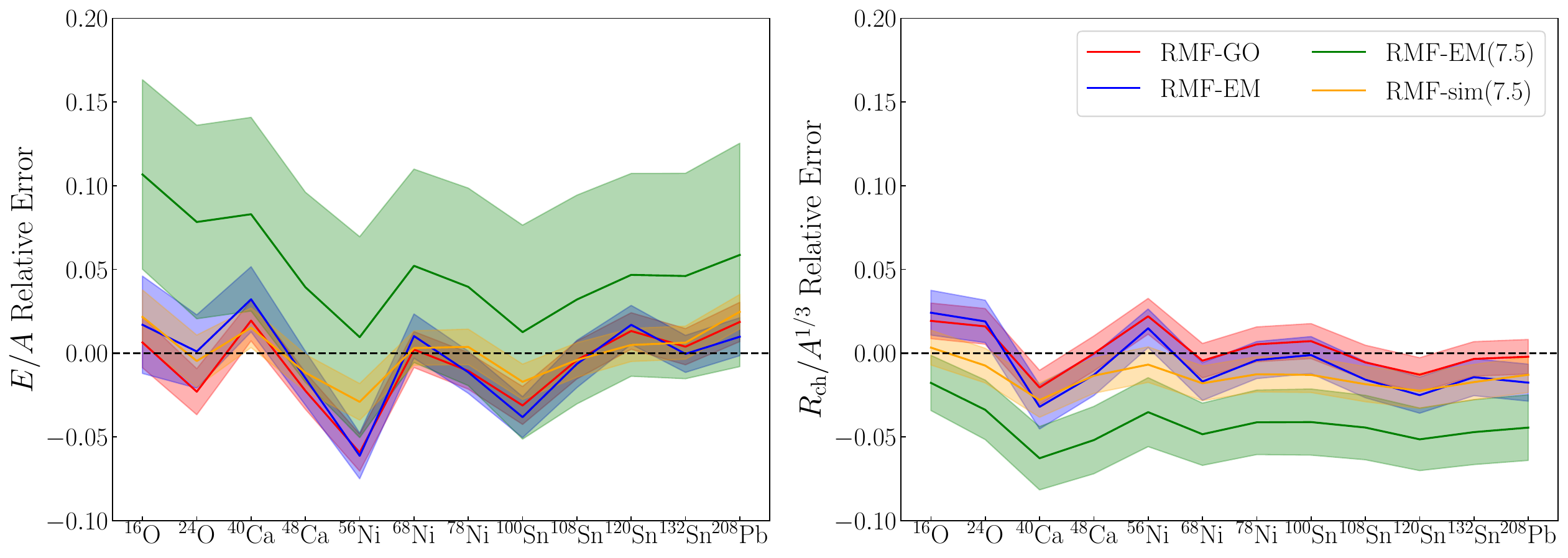}
    \caption{Comparison of relative errors in nuclei for charge radii and energies. The horizontal black line corresponds to zero relative error. Bands show the 68\,\% confidence interval around the mean error. Note that we scale the charge radius by $A^{1/3}$ as in \cref{fig:RMFGO}.}
    \label{fig:3N-models}
\end{figure*}

\subsection{Fits to other interactions}

We now discuss the remaining fits to three other chiral interactions. 
For each fit, we utilize the same model extensions as in \rmfgo{}. 
Of these three interactions, two are modifications of previously existing interactions: \arthuisA{} and \arthuisB{} both modify the 3N couplings $c_D$ and $c_E$. 
When fitting the RMF model to these two interactions, we find considerable difficulty in reproducing many properties of nuclei when compared to fits to the \dnnlogo{} and \hebelerA{} interactions.
This can be seen in \cref{fig:3N-models} where we show the relative errors of each RMF model to the energies and charge radii predicted in \textit{ab initio} calculations. 
Although both \rmfgo{} and RMF-EM show similar error trends and similar error sizes, the other two interactions show considerably different trends and, in the case of RMF-EM(7.5), much larger errors.
This inability for the RMF models to accurately capture the effects of the stronger short-range $3N$ forces is potentially indicative of missing interactions within the RMF models we consider. 
The particularly large uncertainties for the RMF-EM(7.5) model are due to the relatively larger nuclear matter uncertainties in the \textit{ab initio} input data, highlighting the importance of nuclear matter constraints.

\begin{figure}
    \centering
    \includegraphics[width=\columnwidth]{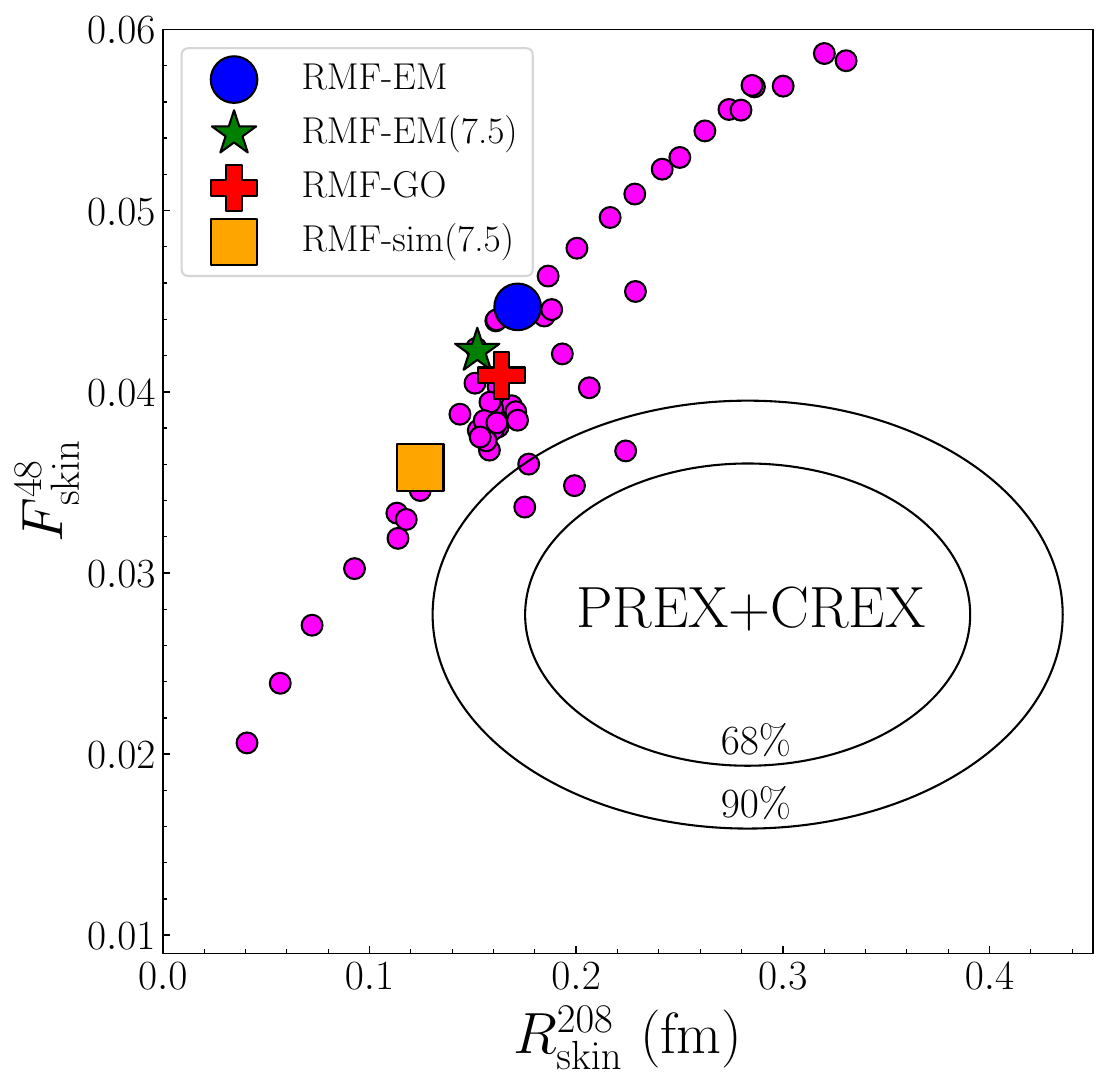}
    \caption{Comparison of the neutron skin ($R_n-R_p$) of \elem{Pb}{208} to the form factor skin ($F_{\rm ch}-F_{\rm wk}$) in \elem{Ca}{48} as determined by the PREX-1+2 \cite{PREX,PREX2} and CREX \cite{CREX} experiments. The ellipses show the 68\% and 90\% confidence intervals determined from both experiments.}
    \label{fig:PREXCREX}
\end{figure}

We provide the full list of Lagrangian couplings in \cref{tab:yukawa_couplings,tab:nonlinear_couplings} as well as the nuclear matter properties for each model in \cref{tab:bulk_props}. 
Additionally, we show the comparison of all the models presented here to the results of the PREX and CREX experiments in \cref{fig:PREXCREX}.
We plot the neutron skin of \lead\ against the form factor skin ($F_{\rm ch}-F_{\rm wk}$) of \calfe\ with magenta points showing a sample of RMF models from Ref.~\cite{Reed:2023}. 
In \cref{fig:PREXCREX}, none of the models shown in this work lie within the 90\,\% confidence ellipse. 
This is a known issue within density functional theory generally, although some works have suggested that a strong spin-orbit force may allow for reconciliation of the two experimental results \cite{PREX:2021umo,CREX}.
The interactions from chiral EFT used in this work do not lead to a large neutron skin in \lead, consistent with \textit{ab initio} calculations for the neutron skin of \elem{Pb}{208}~\cite{Hu2022NP_Pb208, Arthuis2024arxiv_LowResForces}.

\begin{table*}
    \renewcommand{\arraystretch}{1.3}
    \begin{ruledtabular}
    \begin{tabular}{l c c c c c }
    Model & $m_s$ & $g_s^2$ & $g_v^2$ & $g_\rho^2$ & $g_\delta^2$ \\\hline\hline
    \rmfgonodelta{}& $ 466.9709\pm0.1073 $ & $101.0793\pm0.4023$ & $ 197.7524\pm0.0349 $ & $ 139.4445\pm0.1150 $ & 0.0  \\\hline
    \rmfgo{}& $ 523.2300\pm1.8179 $ & $ 129.8895\pm1.3629 $ & $ 207.3669\pm0.4664 $ & $ 187.2445\pm2.8144 $ & $ 87.2318\pm1.4154 $ \\\hline
    RMF-EM & $ 512.0006\pm0.0185 $ & $ 119.4313\pm0.3547 $ & $ 184.6073\pm0.0400 $ & $ 194.0101\pm0.1319 $ & $ 96.3565\pm0.0339 $ \\\hline
    RMF-EM(7.5) & $ 502.1187\pm0.2811 $ & $ 120.7120\pm0.6200 $ & $ 194.0022\pm0.0132 $ & $ 218.3965\pm0.1114 $ & $ 92.8285\pm0.0266 $ \\\hline
    RMF-sim(7.5) & $ 534.5869\pm0.9979 $ & $ 132.5964\pm0.7630 $ & $ 203.1945\pm0.4788 $ & $ 225.3631\pm0.3318 $ & $ 91.1643\pm0.2010 $
    \end{tabular}
    \end{ruledtabular}
    \caption{Yukawa sector meson field couplings for the various RMF models discussed in the text. This is supplemented by the nonlinear and tensor couplings found in \cref{tab:nonlinear_couplings}. Note that the scalar meson mass $m_s$ is given in units of MeV. We fix the masses of the other mesons to their experimental values of $m_v=782.5$ MeV, $m_\rho=763$ MeV, and $m_\delta=980$ MeV \cite{PDG2020}.}
    \label{tab:yukawa_couplings}
\end{table*}

\begin{table*}
    \renewcommand{\arraystretch}{1.3}
    \begin{ruledtabular}
    \begin{tabular}{l c c c c c c}
    Model & $\kappa$ & $\lambda$ & $\zeta$ & $\Lambda_v$ &   $f_v$ & $f_\rho$ \\\hline\hline
    \rmfgonodelta{} & $ 2.8198\pm0.0675 $ & $ 0.0136\pm0.0102 $ & $ 0.0605\pm0.0238 $ & $ 0.0503\pm0.0019 $ & 0.0 & 0.0 \\\hline
    \rmfgo{} & $ 2.0458\pm0.2901 $ & $ 0.0245\pm0.0059 $ & $ 0.0756\pm0.0141 $ & $ 0.0198\pm0.0007 $  &  $ 11.6757\pm1.0396 $ & $ -107.0218\pm6.7774 $ \\\hline
    RMF-EM & $ 5.0335\pm0.1445 $ & $ 0.0098\pm0.0167 $ & $ 0.0897\pm0.0436 $ & $ 0.0208\pm0.0018 $ & $ 13.7078\pm0.1603 $ & $ -102.9553\pm0.0427 $ \\\hline
    RMF-EM(7.5) & $ 4.6617\pm0.0683 $ & $ 0.0276\pm0.0126 $ & $ 0.1204\pm0.0359 $ & $ 0.0260\pm0.0017 $ & $ 10.7080\pm0.0965 $ & $\hphantom{1}$$-55.2602\pm0.0182 $ \\\hline
    RMF-sim(7.5) & $ 1.8957\pm0.1642 $ & $ 0.0207\pm0.0064 $ & $ 0.0642\pm0.0131 $ & $ 0.0237\pm0.0009 $ & $ 11.5402\pm0.8702 $ & $ -117.4315\pm0.2482 $ 
    \end{tabular}
    \end{ruledtabular}
    \caption{Nonlinear and tensor couplings for the various RMF models described in the text. All couplings are dimensionless except $\kappa$ which has units of MeV.}
    \label{tab:nonlinear_couplings}
\end{table*}

\begin{table*}
    \renewcommand{\arraystretch}{1.3}
    \begin{ruledtabular}
\begin{tabular}{l c c c c c c }
Model & $\rho_0$ & $E/A$ & $K_0$ & $J$ & $L$ & $K_{\rm sym}$ \\
& (fm$^{-3}$) & (MeV) & (MeV) & (MeV) & (MeV) & (MeV) \\\hline\hline
\rmfgonodelta{} & $ 0.1694 \pm 0.0013 $ & $ -16.75 \pm 0.10 $ & $ 211.4 \pm 12.84 $ & $ 31.24 \pm 0.23 $ & $ 58.88 \pm 0.81 $ & $ -14.71 \pm 8.98 $ \\\hline
\rmfgo{} & $ 0.1659 \pm 0.0015 $ & $ -16.75 \pm 0.08 $ & $ 219.2 \pm 8.350 $ & $ 30.77 \pm 0.24 $ & $ 50.80 \pm 0.75 $ & $ -16.32 \pm 7.47 $ \\\hline
RMF-EM & $ 0.2101 \pm 0.0024 $ & $ -17.41 \pm 0.11 $ & $ 208.8 \pm 13.41 $ & $ 34.91 \pm 0.31 $ & $ 60.33 \pm 2.47 $ & $ 22.74 \pm 30.87 $ \\\hline
RMF-EM(7.5) & $ 0.1871 \pm 0.0082 $ & $ -16.99 \pm 0.77 $ & $ 175.0 \pm 13.60 $ & $ 32.25 \pm 1.37 $ & $ 47.58 \pm 3.76 $ & $ 0.763 \pm 5.07 $ \\\hline
RMF-sim(7.5) & $ 0.1626 \pm 0.0018 $ & $ -16.71 \pm 0.09 $ & $ 230.4 \pm 10.31 $ & $ 28.99 \pm 0.25 $ & $ 39.58 \pm 0.99 $ & $ 21.93 \pm 10.72 $ 
\end{tabular}
    \end{ruledtabular}
    \caption{
    \label{tab:bulk_props}
    Summary of bulk nuclear matter properties for each of the RMF models described in the text.}
\end{table*}

\subsection{Neutron star properties}

In this section, we detail the predictions for neutron star properties of the five RMF models presented in the text, \rmfgonodelta{} without the model extensions and \rmfgo{}, RMF-EM, RMF-EM(7.5), and RMF-sim(7.5) with the model extensions. 
These calculations illustrate the importance of high-density couplings within RMF theory, such as the quartic vector self-coupling $\zeta$, and the effects of the $\delta$~meson in the neutron star EOSs. 
For the core of the neutron star, we include electrons and muons when calculating beta equilibrium and charge equilibrium.
The core-crust transition density is calculated using the random-phase approximation method of Ref.~\cite{Carrier:2003} modified to account for the new inclusion of the $\delta$ meson. 
For the outer crust, we use the Duflo-Zuker~\cite{Duflo:1994} mass table up to neutron drip density. 
The inner crust is then found by interpolating between the outer crust and core using a cubic polynomial, ensuring both energy density and sound speed are continuous at each interface as in Ref.~\cite{piekarewicz:2019}. 

We list some important neutron star properties in \cref{tab:NSs}. 
We find that all five models do not reach a maximum mass of 2$\msun$ because of the higher values for $\zeta$ in each model.
Typically, $\zeta$ is tuned to lower values such that the maximum neutron star mass reaches at least 2$\msun$. 
However, we found that fixing this parameter to low values greatly impacted the fits to the SNM EOS away from saturation density. 
Hence, we kept this parameter free to best capture all features present in the nuclear structure and EOS data. 
This discrepancy, that the maximally observed massive neutron star of ~2$\msun$~\cite{Cromartie:2019kug,Fonseca:2021wxt} cannot be reproduced with FSUGold-like models when $\zeta$ is not tuned, has been discussed in many previous works~\cite{FSUGold,Fattoyev:2010mx,FSUGold2}.
As expected with such soft EOSs (due to $\zeta$ being large), the radii and tidal deformabilities are all well within the range consistent with GW170817~\cite{Abbott:PRL2017,Abbott:2018exr}. 
Interestingly, we see that the core-crust transition density $\rho_C$ is pushed to higher densities in the four models with the $\delta$~meson. 
This is indicative that the presence of the $\delta$ meson, keeping SNM and PNM fixed as in the case of \rmfgonodelta{} and \rmfgo{}, significantly impacts the transition density, providing a novel way to increase the size of the neutron star crust while keeping bulk nuclear properties the same. 
Additionally, we also see that the Urca threshold density is achieved at around 4 times saturation for each of the five models, although this quantity is more sensitive to the high-density EOS.

\begin{table}
    \renewcommand{\arraystretch}{1.3}
    \begin{ruledtabular}
\begin{tabular}{l c c c c c c }
Model & $\rho_C$ & $\rho_{\rm Urca}$ & $M_{\rm Urca}$ & $R_{1.4}$ & $\Lambda_{1.4}$ & $M_{\rm max}$ \\
& (fm$^{-3}$)& (fm$^{-3}$) & ($\msun$) & (km) & & ($\msun$) \\\hline
\rmfgonodelta{} & 0.0783 & 0.6405 & 1.51 & 12.06 & 325 & 1.71 \\
\rmfgo{} & 0.0880 & 0.6164 & 1.43 & 11.92 & 296 & 1.66 \\
RMF-EM & 0.1049 & 0.5956 & 1.22 & 11.08 & 174 & 1.58 \\
RMF-EM(7.5) & 0.0892 & 0.7802 & 1.31 & 10.60 & 122 & 1.48 \\
RMF-sim(7.5) & 0.0938 & 0.7452 & 1.63 & 11.91 & 334 & 1.73
\end{tabular}
    \end{ruledtabular}
\caption{Overview of selected neutron star properties as predicted by the five models discussed in the text. We show the core-crust transition density ($\rho_C$) as determined by the RPA method of \cite{Carrier:2003}, the Urca threshold density ($\rho_{\rm Urca})$, and the mass of a neutron star with that central density ($M_{\rm Urca}$), the radius and tidal deformability of a $1.4\msun$ neutron star ($R_{1.4}$, $\Lambda_{1.4}$), and the maximum mass ($M_{\rm max}$).}
\label{tab:NSs}
\end{table}

\section{Discussion}
\label{sec:discussion}

A key motivation of this study is the exploration of how RMF models are able to reproduce the results of \textit{ab initio} calculations using chiral EFT interactions.
We find that RMF models are not fully able to reproduce all trends in nuclei based on fits to nuclear matter and selected closed-shell nuclei.
This is more-so the case for the RMF-sim(7.5) and especially the RMF-EM(7.5) models, which are fit to the chiral EFT results using the \arthuisB{} and \arthuisA{} interactions, respectively.
For these Hamiltonians, the short-range $3N$ forces were fit to yield a good simultaneous reproduction of the \elem{O}{16} binding energy and charge radius, an adjustment that yielded a good overall reproduction of binding energies and charge radii up to \elem{Pb}{208}~\cite{Arthuis2024arxiv_LowResForces}.
The resulting 3N force parameter $c_D=7.5$ is relatively large with respect to conventional chiral EFT interactions including the \hebelerA{} and \dnnlogo{} Hamiltonians.
This gives both the modified trends in predictions for nuclear structure observables (notably charge radii), but also increases the assessed uncertainties for the predictions for nuclei and nuclear matter (stemming from the use of approximate, but systematically improvable many-body methods).
Our RMF models clearly struggle to fully reproduce these modified trends.
We understand this to be partially due to the significant short-range $3N$ forces in these chiral EFT Hamiltonians, which have no analog in our RMF models, and partially due to the larger uncertainties from the input data.

Because of the improvements to microscopic calculations of nuclear systems, it is essential that we utilize their predictive powers in refining functionals within EDF theory. 
While none of the three additional terms discussed within this paper are new, they are only infrequently included. 
Unlike Skyrme-type DFTs, in RMF theories it is more difficult to disentangle the contributions of individual couplings to various observables.
This makes the calibration of RMF models challenging \cite{Fattoyev:2010mx,FSUGold2,Chen:2015,Giuliani2023FP_DFTEC}.
Typically, a systematic study of couplings can help indicate the sensitivities of various observables to individual couplings and identify redundant couplings that can be removed from the Lagrangian, allowing for unambiguous fits to data.
For example, the $\sigma$--$\rho$ cross-coupling was originally included to provide an additional isovector-dependent coupling to mix the short-range scalar meson with the effects of the $\rho$ meson.
However, its effect was shown to be much smaller than the $\omega$--$\rho$ cross-coupling in softening the density-dependence of the symmetry energy~\cite{Horowitz:2000xj,Carrier:2003}.
For this reason, higher-order meson interactions beyond what is presented here have largely been omitted.

Our results indicate that the inclusion of the $\delta$~meson alone improves the isovector properties of the nuclear EOS. 
The impact of the $\delta$~meson has been studied previously for finite and infinite systems \cite{Singh:2014} yielding notable changes to the spin-orbit effects and binding energies. 
In a previous work~\cite{Reed:2023}, the $\delta$~meson's inclusion was essential in reconciling the neutron skins of \calfe\ and \lead\ as it allowed for systematic decreasing of the neutron skin in \calfe\ while keeping $L$ constant. 
This effect, while not quite known at the time, is because the large values for $g_\delta$ cause a large change to spin-orbit effects, which affect the nuclear surface. 
\calfe\ has a large surface energy compared to \lead, which is why its effect was so pronounced. 

The saturation points predicted from chiral EFT Hamiltonians and RMF models show different trends, with RMF models typically obtaining saturation densities of around 0.15 fm$^{-3}$ and results from chiral EFT interactions generally lead to larger saturation densities.
Thus, also the EFT-informed RMF models considered here, have saturation densities larger than 0.15 fm$^{-3}$, which also impacts the results for finite nuclei.
This effect also may help explain the errors seen in the charge radius of \elem{Ca}{40} and $\elem{Ni}{56}$ wherein the spin-orbit forces arising from the model extensions are less impactful compared to more neutron-rich isotopes \cite{Horowitz:2012we}, leaving the bulk of the effects seen arising from the symmetric matter EOS behavior.
Nonetheless, if our RMF Lagrangian were sufficiently general, the effect of having a larger saturation density should be able to be absorbed into the model without sacrificing the reproduction of finite nuclei observables based on the same chiral EFT Hamiltonian.

The high-density component of the nonlinear RMF Lagrangian is largely controlled by the $\zeta$ coupling with only some information coming from $\Lambda_v$~\cite{Fattoyev:2010rx}. 
This sector is demonstrably ineffective at reproducing a maximum-mass neutron star consistent with the heaviest pulsars observed without drastically increasing the predicted radius of a $1.4\msun$ neutron star. 
Typically, $\zeta$ can be chosen arbitrarily to reproduce the maximum-mass neutron star of choice, and then the other couplings are refit to accommodate its chosen value. 
Here, we find that $\zeta$ is also important for determining the behavior of low-density nuclear matter due to the effective $\omega$ mass being large in symmetric nuclear matter at low densities~\cite{Glendenning:2000,FSUGold2}. 
As a result, arbitrary choices of $\zeta$ may lead to disagreement with chiral EFT calculations at low densities in matter while not impacting nuclei very much.
The quartic $\rho$~meson self-coupling $\xi$ \cite{Horowitz:2001ya} extends to the high-density EOS regime without impacting symmetric matter or finite nuclei (except for very large values of $\xi$). 
This term may want to be explored in the future along with other high-density terms, which one may adjust to obtain the desired properties of neutron stars. 
It is also possible that there may exist a phase transition at higher densities that reconciles this issue with $\zeta$ and the maximum neutron-star mass.
More generally, this shows the challenges of general high-density EOS extensions within RMF theory.

We conclude with some possible extensions that could be added to the RMF Lagrangian to further improve the theory. 
The lack of direct \nnn{} interactions is a clear omission from the nonlinear RMF model.
One way to address this is to utilize density-dependent couplings, wherein the \nnn{} effects are included through introducing a density-dependency in the Yukawa coupling for the meson fields~\cite{Serot:1997xg,Typel:1999yq,Lalazissis:2005de,LONG:2006150}. 
This approach to RMF theory omits the nonlinear potential terms and the tensor couplings, but has also been shown to be powerful in both connecting to microscopic interactions and being able to reproduce chiral EFT calculations of PNM.
One possible approach would be to introduce some simple density dependence as in Ref.~\cite{Shen:2010pu}.

It is also possible that additional spin-orbit terms could improve the isoscalar sector in nuclei. 
For example, a scalar spin-orbit Lagrangian term that includes derivatives of the scalar and $\delta$ fields,
\begin{eqnarray}
    \mathcal{L}_{sso}=\bar{\psi}[\gamma^\mu\partial_\mu(f_s\phi + f_\delta(\bm{\tau}\cdot\bm{\delta}))]\psi\,,
\end{eqnarray}
reduces to a form similar to the vector spin-orbit terms found in nonrelativistic Skyrme functionals \cite{Salinas_thesis,Yue:2024}. 
These terms are largely unexplored and could have a noticeable effect on isoscalar observables due to the scalar nature of these mesons. 
Finally, connections to tensor forces within chiral EFT are difficult to obtain within mean-field theories due to their vanishing contribution after the mean-field approximation is taken. 
It is possible that the effects of tensor ($J=2$) mesons could be interesting to study and are a natural bridge to connect tensor forces to RMF theories \cite{tensor_mesons}.

Finally, we mention that relativistic microscopic approaches such as relativistic Brueckner-Hartree-Fock calculations are also available (see Ref.~\cite{SHEN2019103713} and references therein). This has led to successful studies for light nuclei and nuclear matter (see, e.g., Refs.~\cite{shen:2017,Zou:2024}). However, our approach employs nuclear-structure information from heavier nuclei than what can currently be calculated with covariant many-body methods based on covariant chiral interactions. A possible follow-up to this work could use microscopic calculations from covariant interaction models once the systematic uncertainties in corresponding many-body calculations are comparable to state-of-the-art ab initio calculations.

\section{Conclusions}

We have performed several fits of nonlinear RMF models to results from chiral EFT calculations of nuclei and nuclear matter. 
In doing so, we have obtained and documented several results that suggest that RMF theory does quite well at reproducing the pure neutron and symmetric nuclear matter predictions using chiral EFT interactions. 
Simultaneously fitting to nuclear matter and selected nuclear structure observables, we find that the addition of the delta meson and vector meson tensor couplings improves the reproduction of charge radii and neutron skins in finite nuclei compared to the conventional Lagrangian of FSUGold-like models. 
However, in symmetric doubly-magic nuclei we find the additional terms perform worse in reproducing the energies obtained from chiral EFT. 
We also found that predictions for chiral EFT interactions with \nnn{} forces with relatively larger short-range couplings are more poorly reproduced compared to interactions with smaller short-range \nnn{} couplings. 
We believe that introducing additional terms that directly impact the spin-orbit sector of the RMF Lagrangian as well as inclusion of exchange terms into the finite nucleus calculations could improve the agreement with chiral EFT.
Additionally, density-dependent couplings could help capture the effects of \nnn{} forces. 
Lastly, we conclude that the high-density sector of the nonlinear RMF interaction needs some refinement as the predictions of the maximum mass neutron star are too small compared to astrophysical observations if $\zeta$ is not constrained by neutron star observables.

\begin{acknowledgments}
We would like to thank Marc Salinas for providing notes on the implementation of the tensor couplings and helpful discussions about possible model extensions. 
We would also like to thank Chuck Horowitz and Jorge Piekarewicz for valuable discussions and insight for this work.  
This work benefited from useful discussions at the Neutron Rich Matter on Heaven and Earth (22r-2a) workshop at the Institute for Nuclear Theory at the University of Washington, Seattle. This work also benefited from discussions at the Third Frontiers in Nuclear Astrophysics Summer School at Ohio University, supported by IReNA under National Science Foundation Grant No. OISE-1927130.

This work was supported by the U.S. Department of Energy through the Los Alamos National Laboratory. 
Los Alamos National Laboratory is operated by Triad National Security, LLC, for the National Nuclear Security Administration of U.S. Department of Energy (Contract No.~89233218CNA000001).
This work was also supported in part by the European Research Council (ERC) under the European Union's Horizon 2020 research and innovation programme (Grant Agreement No.~101020842), by the U.S. Department of Energy, Office of Science, Office of Advanced Scientific Computing Research, Scientific Discovery through Advanced Computing (SciDAC) NUCLEI program,  by the Laboratory Directed Research and Development Program of Los Alamos National Laboratory under project numbers 20230315ER and 20230785PRD1,
and by the Laboratory Directed Research and Development Program of Oak Ridge National Laboratory, managed by UT-Battelle, LLC, for the U.S.\ Department of Energy, and by the European Union under the
Marie Skłodowska-Curie grant agreement No.~101152722. Views and opinions expressed are however those of the author(s) only and do not necessarily reflect those of the
European Union or the European Research Executive Agency (REA). Neither the European Union nor the granting authority can
be held responsible for them.
This research used resources of the Oak Ridge Leadership Computing Facility located at Oak Ridge National Laboratory, which is supported by the Office of Science of the Department of Energy under contract No.~DE-AC05-00OR22725.
The authors gratefully acknowledge the Gauss Centre for Supercomputing e.V.\ (www.gauss-centre.eu) for funding this project by providing computing time through the John von Neumann Institute for Computing (NIC) on the GCS Supercomputer JUWELS at Jülich Supercomputing Centre (JSC).
\end{acknowledgments}

\appendix
\section{Nuclear matter}

Here we describe the equations that are used to calculate the infinite-matter properties of RMF theory using the Lagrangian of the extended RMF models above. From the interaction Lagrangian in \cref{eq:lagrangian}, the meson fields are calculated from their equations of motion at a given proton and neutron density by solving the following set of coupled equations,
\begin{equation}
\begin{split}
    &\frac{m_s^2}{g_s^2}\Phi +\frac{\kappa}{2}\Phi^2+\frac{\lambda}{6}\Phi^3 = \rho_s^p+\rho_s^n\,,\\
    &\frac{m_v^2}{g_v^2}V_0+\frac{\zeta}{6}V_0^3+2\Lambda_vV_0B_0^2=\rho_p+\rho_n\,,\\
    &\frac{m_\rho^2}{g_\rho^2}B_0+2\Lambda_vV_0^2B_0=\rho_p-\rho_n\,,\\
    &\frac{m_\delta^2}{g_\delta^2}D_0= \rho_s^p-\rho_s^n\,,
\end{split}
\end{equation}
where $\rho$ is the baryon density of species $i=(p,n)$, differentiating between proton and neutrons, and $\rho_s$ is the scalar density. Note that here we make the simplification of combining the Yukawa coupling with their respective meson field such that $$(g_s\phi,g_v W_0,g_\rho b_0,g_\delta \delta_0)\longrightarrow (\Phi,V_0,B_0,D_0)\,.$$ The scalar density can be calculated using standard Dirac theory calculations~\cite{Glendenning:2000}:
\begin{eqnarray}
    \rho_s^i = \frac{1}{\pi^2}\int_0^{k_F^i}k^2\frac{M_i^*}{\sqrt{k^2+M_i^{*2}}}dk\,,
\end{eqnarray}
with the effective Dirac mass $M^*$ and Fermi momentum $k_F^i$. The inclusion of the $\delta$~meson splits the proton and neutron Dirac masses such that
\begin{eqnarray}
    M_i^* = M-\Phi\mp\frac{1}{2}D_0\,,
\end{eqnarray}
where protons are ($-$) and neutrons are ($+$). With the above equations, the energy and pressure of nuclear matter can be calculated. The energy density is given by
\begin{align}
    \epsilon(\rho) & = \nonumber\frac{1}{2}\frac{m_s^2}{g_s^2}\Phi^2-\frac{1}{2}\frac{m_v^2}{g_v^2}V_0^2-\frac{1}{2}\frac{m_\rho^2}{g_\rho^2}B_0^2+\frac{1}{2}\frac{m_\delta^2}{g_\delta^2}D_0^2\\
    \nonumber &\quad +\frac{\kappa}{3!}\Phi^3+\frac{\lambda}{4!}\Phi^4-\frac{\zeta}{4!}V_0^4-\Lambda_v V_0^2B_0^2+\rho V_0\\
   &\quad +\frac{1}{2}(\rho_p-\rho_n)B_0+\epsilon_p(\rho)+\epsilon_n(\rho)\,,
\end{align}
where $\epsilon_i(\rho) = \frac{1}{\pi^2}\int_0^{k_F^i}dkk^2\sqrt{k^2+M^{*2}_i}$.

\section{Finite nuclei}

Within the extended Lagrangian presented above, we may derive the equations of motion for the meson fields and baryonic fields using standard practices. We assume spherical symmetry and that all nuclei are in the ground state. This allows us to write the equations of motion for the Hartree system as:

\begin{equation}
\begin{split}
    \Big( m_{\rm s}^2 - \nabla^2 & \Big) \Phi = g_{\rm s}^2 \Big(\rho_s - \frac{1}{2}\kappa\,\Phi^{2} - \frac{1}{6} \lambda \Phi^{3}\Big),  \\
    \Big( m_{\rm v}^2 - \nabla^2 & \Big) V_0 = g_{\rm v}^2 \Big(\rho_{v} - \frac{1}{2M} \frac{f_{\rm v}}{g_{\rm v}}\nabla\rho_t - \frac{\zeta}{6}V_0^{3} - 2 \Lambda_{\rm v} B_0^{2} V_0 \Big), \\
    \Big( m_{\rho}^2 - \nabla^2 & \Big) B_0 = g_{\rho}^2 \Big( \frac{1}{2} \rho_{v,3} - \frac{1}{4M} \frac{f_\rho}{g_\rho} \nabla\rho_{t,3} - 2 \Lambda_{\rm v} V_0^{2} B_0 \Big),\\
    \Big( m_\delta^2 - \nabla^2 & \Big) D_0 = \frac{1}{2} g_\delta^2 \rho_{s,3}\,,\\
    -\nabla^2 & A_0 = e\rho_p\,,\\
    \Big(\frac{d}{dr}+\frac{\kappa}{r}+&\mathcal{T}\Big)g_{n\kappa}-\Big(E+M-\mathcal{S}-\mathcal{V}\Big)f_{n\kappa}=0\,,\\
    \Big(\frac{d}{dr}-\frac{\kappa}{r}-&\mathcal{T}\Big)f_{n\kappa}+\Big(E-M+\mathcal{S}-\mathcal{V}\Big)g_{n\kappa}=0\,,
\label{eq:hartree}
\end{split}
\end{equation}
where the potentials $\mathcal{S, V, T}$ are
\begin{equation}
    \begin{split}
        &\mathcal{S} = \Phi - \frac{\tau_3}{2}D_0\,,\\
        &\mathcal{V} = V_0 + \frac{\tau_3}{2}B_0 + \frac{1+\tau_3}{2} A_0\,,\\
        &\mathcal{T} = \frac{1}{2M}\left(\frac{f_{\rm v}}{g_{\rm v}}\frac{dV_0}{dr}+ \frac{\tau_3}{2} \frac{f_{\rho}}{g_{\rho}}\frac{dB_0}{dr}\right),
    \end{split}
\end{equation}
and the wave functions $g_{n\kappa}$ and $f_{n\kappa}$ are the upper and lower components of a Dirac spinor
\begin{eqnarray}
    \psi = \frac{1}{r}
    \begin{pmatrix}
        ig_{n\kappa}(r)\mathcal{Y}_{\kappa m}(\theta,\phi,s)\\
        -f_{n\kappa}(r)\mathcal{Y}_{-\kappa m}(\theta,\phi,s)
    \end{pmatrix} .
\end{eqnarray}
This allows us to compute the particle densities from the Dirac wave functions
\begin{equation}
    \begin{split}
        &\rho_s(r) = \sum_{\alpha=1}^N \frac{2J_\alpha+1}{4\pi r^2}\Big[g_{n\kappa}^2(r)-f_{n\kappa}^2(r)\Big],\\
        &\rho_v(r) = \sum_{\alpha=1}^N \frac{2J_\alpha+1}{4\pi r^2}\Big[g_{n\kappa}^2(r)+f_{n\kappa}^2(r)\Big],\\
        &\rho_{v,3}(r) = \sum_{\alpha=1}^N \frac{2J_\alpha+1}{4\pi r^2}\Big[g_{n\kappa}^2(r)+f_{n\kappa}^2(r)\Big]\frac{\tau_3}{2},\\
        &\rho_{s,3}(r) = \sum_{\alpha=1}^N \frac{2J_\alpha+1}{4\pi r^2}\Big[g_{n\kappa}^2(r)-f_{n\kappa}^2(r)\Big]\frac{\tau_3}{2},\\
        &\rho_{t}(r) = \sum_{\alpha=1}^N \frac{2J_\alpha+1}{4\pi r^2}\Big[2g_{n\kappa}(r)f_{n\kappa}(r)\Big],\\
        &\rho_{t,3}(r) = \sum_{\alpha=1}^N \frac{2J_\alpha+1}{4\pi r^2}\Big[2g_{n\kappa}(r)f_{n\kappa}(r)\Big]\frac{\tau_3}{2}.
    \end{split}
\end{equation}
where $\alpha$ runs over all occupied shell states in the nucleus ($N$). Using these equations along with standard relativistic Hartree practices \cite{horowitz:1981}, we are able to self-consistently solve for the ground state of spherical closed-shell nuclei.
Charge and weak densities are computed according to Ref.~\cite{Horowitz:2012we}.

\begin{figure*}[p!]
    \centering
    \includegraphics[width=\linewidth]{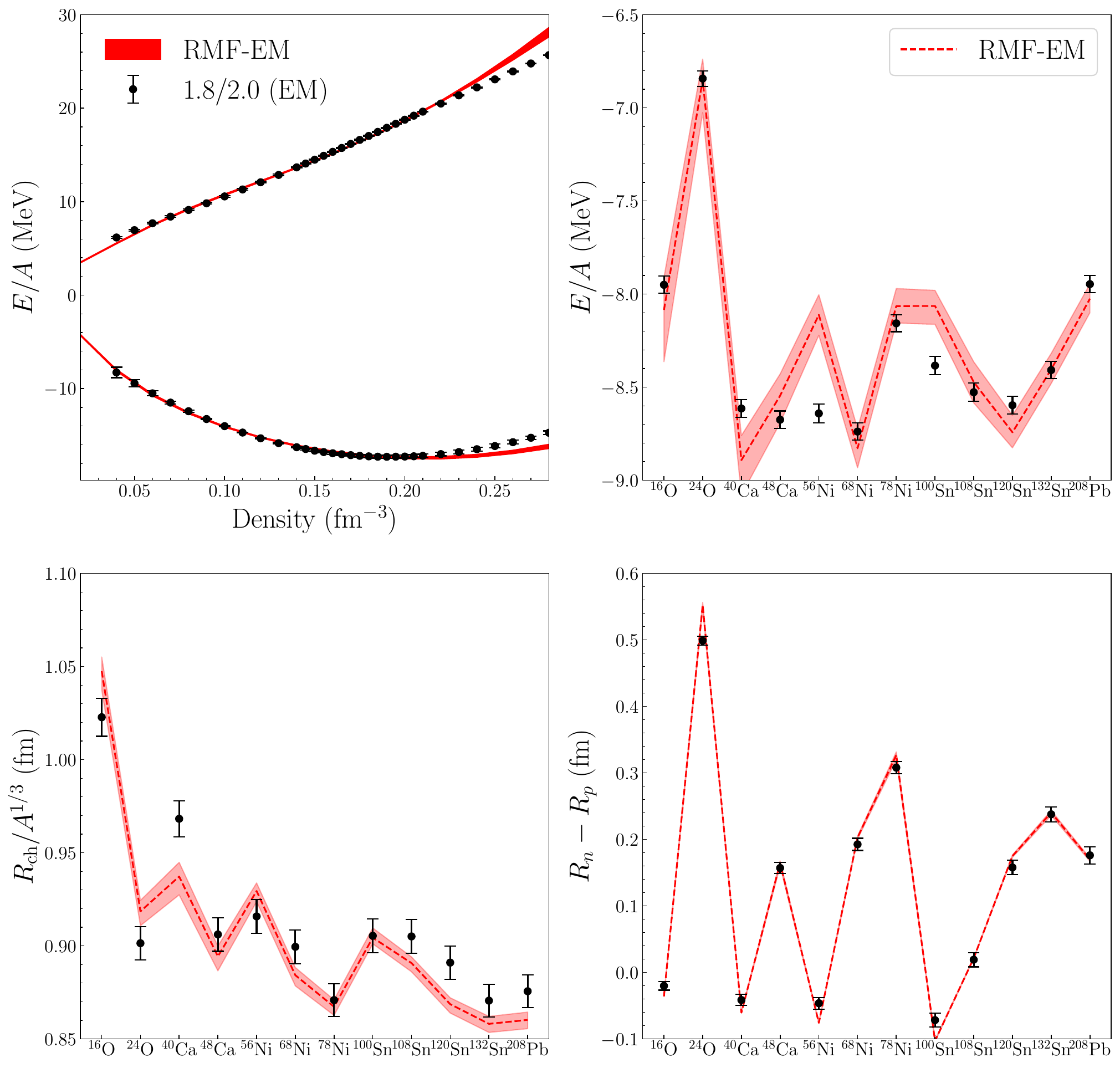}
    \caption{Same as \cref{fig:RMFGO} except for the RMF-EM model.}
    \label{fig:rmf-em}
\end{figure*}

\begin{figure*}[p!]
    \centering
    \includegraphics[width=\linewidth]{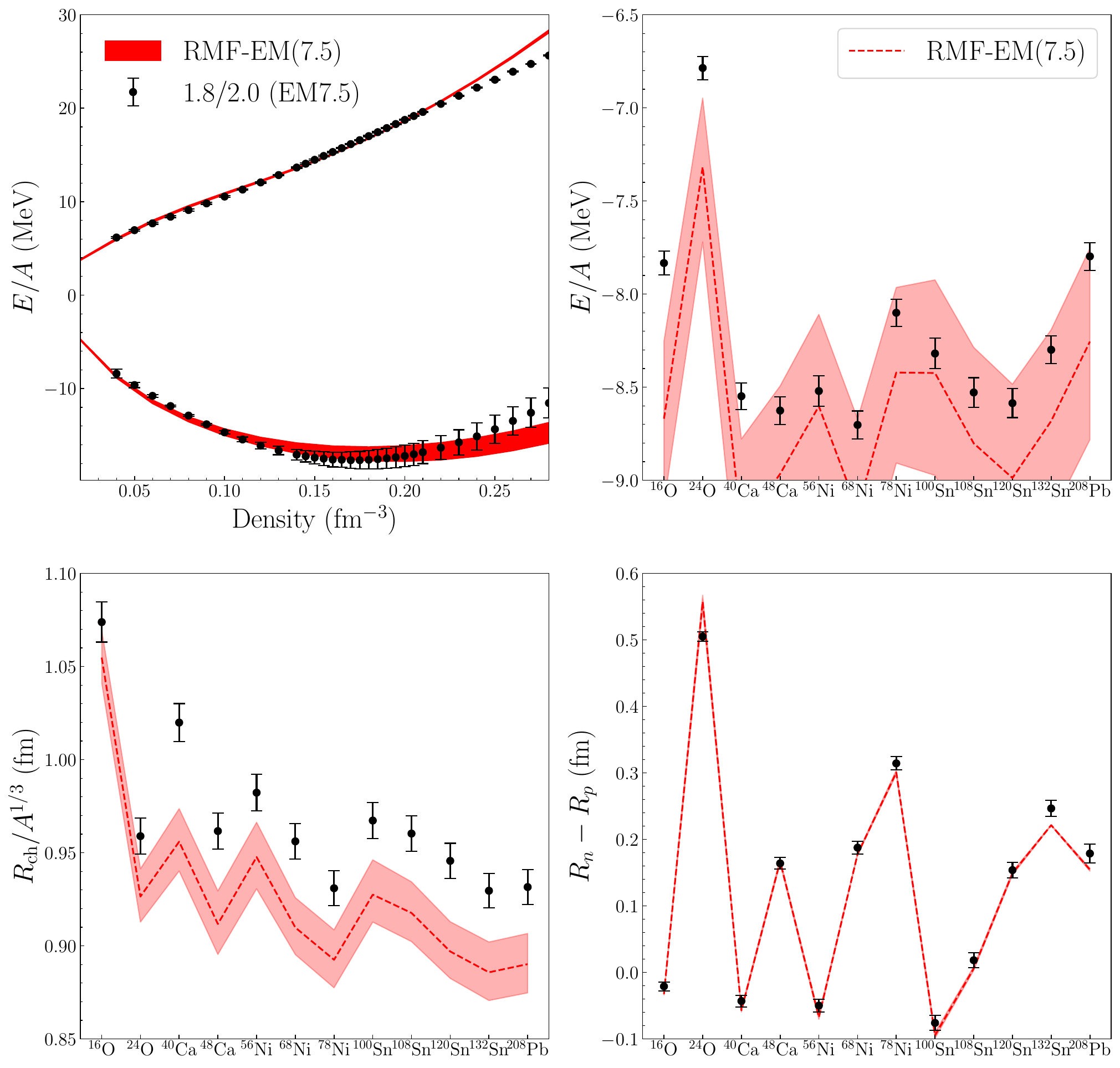}
    \caption{Same as \cref{fig:RMFGO} except for the RMF-EM(7.5) model.}
    \label{fig:rmf-em75}
\end{figure*}

\begin{figure*}[p!]
    \centering
    \includegraphics[width=\linewidth]{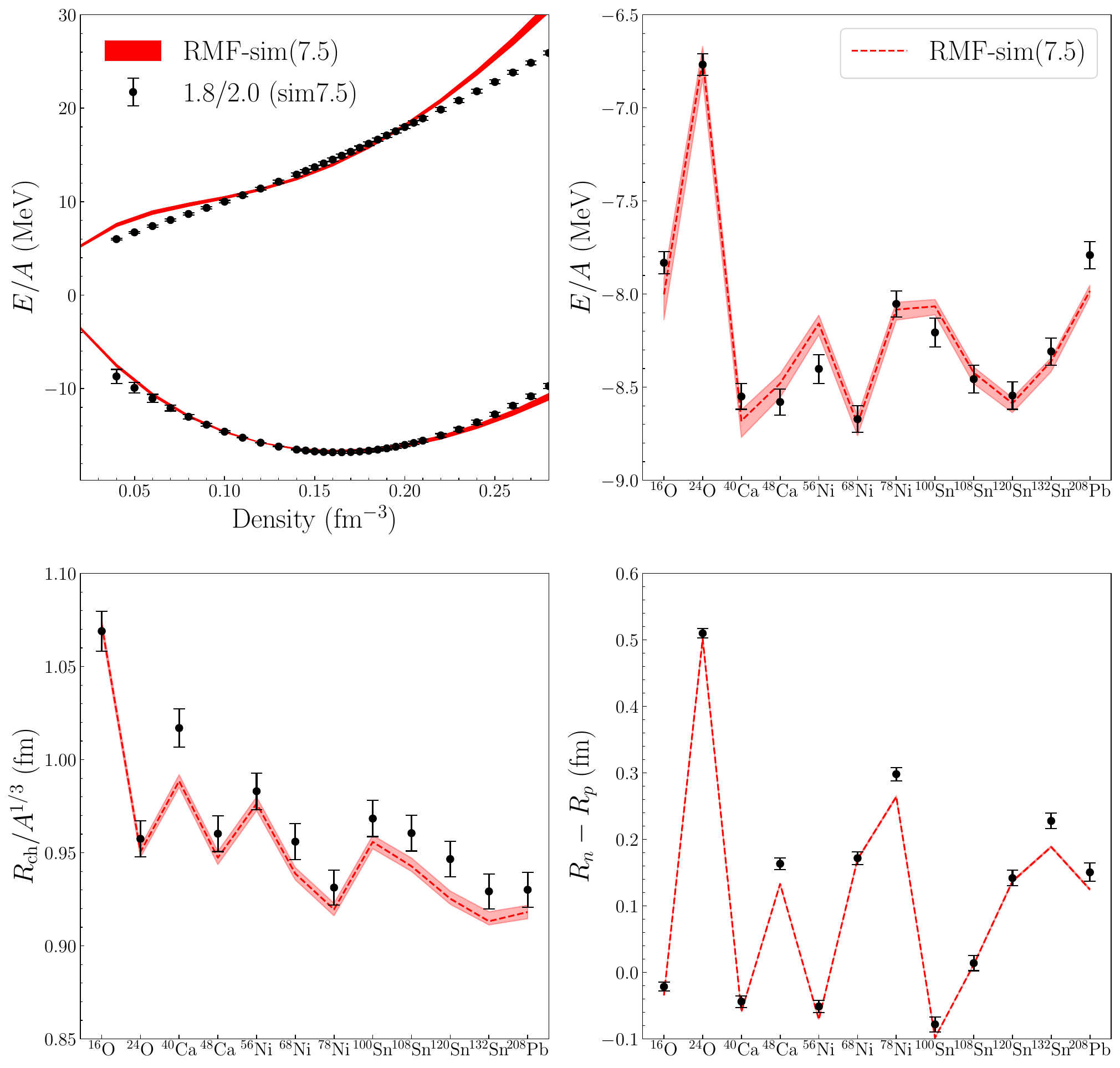}
    \caption{Same as \cref{fig:RMFGO} except for the RMF-sim(7.5) model.}
    \label{fig:rmf-sim75}
\end{figure*}

\section{Additional MCMC sampling of chiral-EFT-informed RMF models}

Here we show the fits from RMF-EM, RMF-EM(7.5), and RMF-sim(7.5) compared to the results from calculations with these chiral EFT interactions~\cite{Arthuis2024arxiv_LowResForces,Alp2025_NuclearMatterMBPT}. Firstly, we see in \cref{fig:rmf-em} (RMF-EM) similar results to \rmfgo{}, where the energies of \elem{Ni}{56} and \elem{Sn}{100} are systematically underbound compared to the neutron-rich nuclei in our selection. Likewise, we see the charge radius of \elem{Ca}{40} is lower than the chiral EFT result, and the neutron skins are well within the chiral EFT predictions. Additionally, we find that the saturation density is determined somewhat at a higher value than what is reported in Ref.~\cite{Alp2025_NuclearMatterMBPT} due to the lack of higher density points included in the fit. 

In \cref{fig:rmf-em75,fig:rmf-sim75}, however, we find considerable difficulties in reproducing the chiral EFT results with our RMF models. In particular, these have stronger short-range 3N forces than in the other chiral EFT interactions used here. In doing so, the SNM EOS has larger error bars than the two previously discussed interactions, which leads to larger uncertainties in the energies and charge radii across the nuclei considered here. 
Three-nucleon forces are not explicitly included in the RMF Lagrangian utilized here, so the effect of this refit is not fully captured in the RMF model.
\clearpage

\bibliography{references}

\end{document}